# Practical one-shot data-driven design of fractional-order PID controller: Fictitious reference signal approach

Ansei Yonezawa #1*, Heisei Yonezawa#1, Shuichi Yahagi#2, Itsuro Kajiwara#1

#1

Division of Mechanical and Aerospace Engineering, Hokkaido University

N13, W8, Kita-ku, Sapporo, Hokkaido 060-8628, Japan

#2

6th Research Department, ISUZU Advanced Engineering Center Ltd.

8 Tsutidana, Fujisawa-shi, Kanagawa 252-0881, Japan

**Declaration of Competing Interest**

The authors declare that they have no known competing financial interests or personal relationships that could have appeared to influence the work reported in this paper.

**Acknowledgment**

This work was supported in part by JSPS KAKENHI Grant Number JP23K19084.

*Corresponding author (e-mail address: ayonezawa[at]eng.hokudai.ac.jp)

# Practical one-shot data-driven design of fractional-order PID controller: Fictitious reference signal approach

**Abstract**

This study proposes a one-shot data-driven tuning method for a fractional-order proportional-integral-derivative (FOPID) controller. The proposed method tunes the FOPID controller in the model-reference control formulation. A loss function is defined to evaluate the match between a given reference model and the closed-loop response while explicitly considering the closed-loop stability. A loss function value is based on the fictitious reference signal computed using the input/output data. Model matching is achieved via loss function minimization. The proposed method is simple and practical: it needs only one-shot input/output data of a plant (no plant model required), considers the bounded-input bounded-output stability of the closed-loop system from a bounded reference input to a bounded output, and automatically determines the appropriate parameter value via optimization. Numerical simulations show that the proposed approach facilitates good control performance, and destabilization can be avoided even if perfect model matching is unachievable.

## 1. Introduction

### 1.1. Motivation

Fractional calculus (FC) is a mathematical framework that considers integrals and derivatives of arbitrary (not necessarily integer) order [1]. FC serves as a natural extension of traditional integer-order (IO) calculus. Owing to its flexibility and ability to richly express various phenomena, FC has attracted considerable attention in recent years across numerous fields of engineering [2]. In the field of system and control engineering, various studies have developed theories of system modeling, analysis, and controller synthesis based on the FC framework, and have examined their effectiveness for real-world applications. In [3], fractional-order (FO) fuzzy proportional-integral-derivative (PID) controller has been employed for pitch control of a wind turbine. FO neural sliding mode control has been developed for piezoelectric actuator [4]. The dielectric elastomer actuator has been modeled as the FO transfer function [5]. In [6], the effectiveness of FC to control continuum soft robots has been discussed. In addition, a novel system identification technique has been proposed for an FO multiple-input single-output system [7], several FO active disturbance rejection controllers have been examined [8,9], and an enhanced FO sliding mode control method has been developed for a class of FO uncertain systems [10].

FOPID control [11] is a typical and practical FO control methodology. The FOPID controller consists of a proportional term, an FO integral term, and an FO derivative term. FOPID control can be regarded as a natural generalization of conventional IOPID control with two additional tuning knobs, namely, the integral and differential orders. Owing to its higher design flexibility, FOPID control is an advanced practical control method, which is expected to replace the conventional IOPID control [12]. In fact, several studies have reported that FOPID control outperforms IOPID control in terms of control performance and robustness for *both* IO and FO plants [13]. For instance, the FOPID controller can provide larger achievable region of the gain crossover frequency and phase margin for first order plus time delay systems than the IOPID controller [14]. Several experimental studies have shown that FOPID control can outperform IOPID control for a permanent magnet synchronous motor modeled as the IO transfer function [15], and for the heat flow system showing the FO property [16]. As demonstrated in previous studies, proper utilization of FO control should significantly improve the safety, reliability, and performance of many practical automatic control systems. Therefore, this study focuses on FO controllers including the FOPID controller. It should be noted that the tuning of proportional, integral, and derivative gains *and* integral and derivative order is an important (perhaps the most important) task for reaping the benefits of FOPID control.

### 1.2. Related work

Compared with the tuning of the IOPID controller, the tuning of the FOPID controller is more challenging owing to the increase in the number of tuning parameters. To overcome this difficulty, many attempts have been made to establish a practical tuning policy of the FOPID controller [17–20]. In [18,19], fruitful tuning schemes have been established for some special plant classes. Graphical approaches have been developed for tuning the FOPID controller: the iso-slope phase curve [17], plots based on frequency response data [20]. However, graphical approaches involve manual trial-and-error, increasing the designers' burden. Several studies have proposed FOPID controller tuning based on the plant model: analysis of the achievable specifications for the flat-phase design method [21,22], utilization of the delayed Bode's ideal transfer function [23,24], time-domain analysis [25], to name a few. Nevertheless, obtaining the plant model is expensive and tedious [26–28]. Moreover, the obtained plant model may not be appropriate for designing a control system [29]. Parameter tuning based on a numerical optimization algorithm is a low-cost and effective approach for tuning the FOPID controller [30]. For example, the genetic algorithm [31], the differential evolution algorithms [32], and the chaotic atom search algorithm [33] have been employed for tuning the FOPID controller. An extensive survey of this approach can be found in [12]. In general, however, this approach also requires a plant model so that control simulations are conducted repeatedly to evaluate a performance index that is to be minimized by an optimization algorithm.

*Remark 1.1*: In general, FO filters are approximated using IO ones for implementing FOPID controllers in practice [12] (various good approximation schemes are available [34]). Moreover, discretization is necessary to implement the controller in a digital computer [35]. Note that even though such an IO approximation and discretization are in order, the FOPID controller can still achieve superior performance compared to the IOPID controllers [36].

Recently, numerous efforts have been made to develop advanced control strategies so as to solve challenging engineering problems. Among them, data-driven control methods have attracted considerable attention [29]. Data-driven control techniques use input/output data for controller design instead of a model of a plant derived via first principles. Data-driven control methods can be classified as *indirect* and *direct* methods [29]. Unlike the indirect method, which first explicitly identifies the plant model using the data, the direct method avoids such system identification from the data and directly designs the control policy from the data [37]. Various direct data-driven controller tuning methods for solving the model-reference control problem have been examined, such as iterative feedback tuning (IFT) [38], correlation-based approach [39], virtual reference feedback tuning (VRFT) [40,41], and fictitious reference iterative tuning (FRIT) [42]. Owing to their usefulness, many studies have applied these methods to various control problems [43–47].

These direct data-driven controller tuning schemes for model-reference control should be effective and practical solutions for tuning of the FOPID controller. This is because they do not require any plant model or identification procedure and can automatically determine the appropriate controller parameters via optimization. However, despite their usefulness, there is no practical direct data-driven tuning method of FO controllers to the best of our knowledge. The FOPID controller has been tuned on the basis of IFT, which involves a large number of experiments [48]. One study tuned the FOPI controller via the VRFT [49]. However, this method requires a frequency response computation so as to estimate the stability region. Another study has also examined the VRFT for FOPI control [50]; only the proportional and integral gains have been tuned via the VRFT. Consequently, a new data-driven tuning technique is required to improve the practicality of the FOPID controller and exploit the advantage of FO control for various systems.

*Remark 1.2*: A major challenge of direct data-driven controller tuning is the consideration of the closed-loop stability [43,51]. Note that in the data-driven FOPID controller tuning, the closed-loop stability should be considered for the IO-approximated and discretized (i.e., ready-to-implement) controller since the evaluation for the ready-to-implement controller is more reliable.

### 1.3. Contribution and novelty of the study

This study proposes a new direct data-driven tuning technique of the FOPID controller for a single-input single-output (SISO) linear time-invariant (LTI) system. This approach requires only one-shot input/output data and achieves the required controlled response and closed-loop stability with high design efficiency and practical applicability. The proposed technique solves the model-reference control problem, i.e., the FOPID controller is tuned so as to achieve the resultant closed-loop property as close to a given reference model as possible. The closed-loop response is evaluated on the basis of a fictitious reference signal. The fictitious reference signal is computed using only one-shot input/output data. The controller tuning for achieving model matching is formulated as the minimization of a loss function defined using the fictitious reference signal and the reference model. The loss function explicitly considers the bounded-input bounded-output

(BIBO) stability of the closed-loop system from a bounded reference input to a bounded output; minimizing the loss function yields the controller parameter for realizing model matching and BIBO stability. Such a consideration is performed for the IO-approximated and discretized controller. The effectiveness of the proposed technique is confirmed via numerical experiments. Consequently, this study facilitates industrial application of FO control by providing a practical implementation technique, contributing the improvement of the safety, reliability, and performance of many practical control systems.

In particular, the contributions and novelties of this study can be summarized as follows:

1) (*Simplicity*) The proposed tuning method follows the direct data-driven control approach: no analytical derivation of plant models, computation of the frequency response, or explicit system identification is required. This is a significant advantage over conventional analytical methods requiring the plant model and the optimization-based approach based on the simulations using the plant model. This tuning scheme requires only a one-shot closed-loop experiment, i.e., repeated experiments are avoided. Therefore, the proposed approach is simpler than previous data-driven methods developed for the FOPID controller (e.g., the IFT-based method involving repeated experiments [48], the VRFT-based approach requiring computation of the frequency response [49]).

2) (*Reduced burden on designers*) After the data is collected, the appropriate parameters of the FOPID controller are automatically determined via an optimization algorithm. Note that in contrast to the previous study [50], the controller parameters determined using the proposed scheme include the integral and derivative orders. The proposed method does not require manual tuning based on the designer's trial-and-error approach. Hence, it significantly reduces the implementation burden.

3) (*Stability consideration*) During the tuning, the BIBO stability of the closed-loop system from a bounded reference input to a bounded output is explicitly considered for the IO-approximated and discretized (i.e., *ready-to-implement*) FOPID controller. This feature yields a practically feasible controller. Note that unlike the existing method [49], the proposed tuning scheme does not require explicit computation of the frequency response so as to consider the stability.

4) (*Practical implementation of FO control*) The proposed method is a practical approach for implementing the FOPID controller owing to the three features discussed above. Moreover, the proposed scheme is applicable not only to the FOPID controller but also to various FO controllers taking the transfer functions. Therefore, this study extends the practical utilization of FO control to on-site applications, thereby bridging the gap between theory and practice of FO control. (In addition, the proposed technique can also be applicable to IO controllers taking the transfer functions, as demonstrated in Section 4).

### 1.4. Structure of the paper

The remainder of this paper is organized as follows. Section 2 provides some preliminaries and states the problem under consideration. Section 3 describes the proposed method. Section 4 presents numerical examples to demonstrate the validity of the proposed method and compares the performances of the FOPID and IOPID controllers. Section 5 discusses the results. Finally, Section 6 concludes the paper.

## 2. Preliminaries

### 2.1. Notation

The symbols $\mathbb{R}$, $\mathbb{R}_+$, and $\mathbb{R}^n$ are the sets of real numbers, strictly positive real numbers, and $n$-dimensional real vectors, respectively. The Laplace variable and the $Z$-variable are denoted by $s$ and $z$, respectively. For a continuous-time transfer function $G(s)$, $G(s;\theta)$ denotes $G(s)$ with tunable parameters $\theta \in \mathbb{R}^n$ (the discrete-time case is represented in the same manner). Further, c2d($G(s)$) denotes the discretization of $G(s)$. For a continuous-time FO transfer function $G_{FO}(s)$, its IO approximation via some approximation method is denoted as F2I($G_{FO}(s)$).

For an input $u = \{u_k\}_{k=0}^{\infty}$, the output $y = \{y_k\}_{k=0}^{\infty}$ of a discrete-time IO proper rational SISO LTI system $G(z)$ can be computed as $y_k = \sum_{\tau=0}^{k} g_\tau u_{k-\tau}$, where $g_\tau$ is the impulse response of $G(z)$ at time $\tau$. We sometimes denote it as $y_k = G(z)u_k$, and, similarly, $y = G(z)u$.

Let $v = [v_1 \ v_2 \ \cdots \ v_n]^T \in \mathbb{R}^n$ be a real-valued vector. The standard $p$-norm of $v$ is denoted by $\|v\|_p$, i.e., $\|v\|_p := \{\sum_{i=1}^{n}|v_i|^p\}^{\frac{1}{p}}$ for $1 \le p < \infty$ and $\|v\|_p := \max_i |v_i|$ for $p = \infty$. Similarly, for a real-valued signal $u = \{u_k\}_{k=0}^{\infty}$, $\|u\|_p$, the $\ell_p$-norm of $\{u_k\}_{k=0}^{\infty}$, is defined conventionally as $\|u\|_p := \{\sum_{k=1}^{\infty}|u_k|^p\}^{\frac{1}{p}}$ for $1 \le p < \infty$ and $\|u\|_p := \sup_k |u_k|$ for $p = \infty$.

### 2.2. FOPID controller and its IO-approximation and discretization

The continuous-time FOPID controller $C_{FOPID}$ can be expressed as follows [11]:

$$C_{FOPID}(s;\theta) = K_{fp} + K_{fi}s^{-\lambda} + K_{fd}s^{\mu}, \tag{1}$$

where $\theta := [K_{fp} \ K_{fi} \ \lambda \ K_{fd} \ \mu]^T \in \mathbb{R}^5$ denotes tuning parameters that are to be determined by the proposed method. Further, $\lambda, \mu \in \mathbb{R}_+$ are integral and derivative orders, respectively. Note that $\lambda$ and $\mu$ do not necessarily take the integer values; if $\lambda = \mu = 1$, (1) becomes the traditional continuous-time IOPID controller.

The FO integral and derivative are defined by improper integrals or infinite summation [1]. Therefore, $s^{-\lambda}$ and $s^{\mu}$, i.e., the FO integral and derivative operators, must be IO-approximated for practical implementations [13]. Various approximation schemes based on an IO transfer function are available [34]. Then, the IO-approximated transfer function is discretized (e.g., the Tustin transform) to implement the FO filter in a digital computer.

### 2.3. BIBO stability for IO proper rational SISO LTI system

Consider an IO proper rational SISO LTI transfer function $G(z)$ with the impulse response $g_k$ at discrete time $k$. Then, the following are standard.

*Definition 2.1*: $G(z)$ is said to be BIBO stable if $G(z)$ satisfies $\|G(z)u\|_\infty < \infty$ for any input $u = \{u_k\}_{k=0}^\infty$ satisfying $\|u\|_\infty < \infty$.

*Lemma 2.1*: $G(z)$ is BIBO stable if $\sum_{k=0}^\infty |g_k| < \infty$.

(*Proof*) It is clear from $\|G(z)u\|_\infty < \|u\|_\infty \{\sum_{k=0}^\infty |g_k|\}$. ∎

### 2.4. Problem formulation

Let IO rational SISO LTI transfer functions $P(z)$ and $M_D(z)$ be a plant to be controlled and the reference model, respectively, and $C(z;\theta) := \text{c2d}(\text{F2I}(C_{FOPID}(s;\theta)))$ be an IO-approximated discrete-time FOPID controller with parameter $\theta$. $M_D(z)$ is a given by a designer, which specifies the desired response of the closed-loop system. Here, $M_D(z)$ is designed to be proper and BIBO stable. In this study, we consider the direct data-driven controller tuning in the model-reference control problem as shown in Fig. 1, where $r$, $u$, and $y$ are the set point reference signal, control input, and output, respectively, and the model of $P(z)$ is unknown. The reference $r$ is assumed to be bounded.

*Assumption 2.1*: $P(z)$ is proper. $C(z;\theta)$ is proper for any $\theta \in \Theta$, where $\Theta$ is the search range of $\theta$.

*Remark 2.1*: As long as Assumption 2.1 is satisfied, various IO approximation methods and discretization methods can be adopted. For example, the Oustaloup recursive filter [52] together with the Tustin discretization can be used (perhaps this is the most common approximation method).

Now, suppose that $u_{0,[0,N]} = \{u_{0,k}\}_{k=0}^N$ and $y_{0,[0,N]} = \{y_{0,k}\}_{k=0}^N$, i.e., the input and output data, are obtained via a closed-loop experiment using $C(z;\theta_0)$, i.e., a controller where the tuning parameters are set as $\theta_0$, and a reference signal $r_{0,[0,N]} = \{r_{0,k}\}_{k=0}^N$. The following assumptions on $r_{0,[0,N]}$ are in order.

*Assumption 2.2*: $r_{0,0} \neq 0$ and $\|r_{0,[0,N]}\|_\infty < \infty$.

Here, let $T(z;\theta)$ be the closed-loop transfer function from $r$ to $y$ given by $C(z;\theta)$ in Fig. 1, i.e., $T(z;\theta) := P(z)C(z;\theta)\{1 + P(z)C(z;\theta)\}^{-1}$. Then, the direct data-driven model-reference control problem in this study is formulated as follows:

*Problem 2.1*: Provided that the plant model $P(z)$ is unknown, find the optimal parameter $\theta^* \in \mathbb{R}^5$ of the FOPID controller using $u_{0,[0,N]}$, $y_{0,[0,N]}$, $r_{0,[0,N]}$, and $M_D(z)$, such that $T(z;\theta^*)r_{0,[0,N]}$ is as close to $M_D(z)r_{0,[0,N]}$ as possible.

*Remark 2.2*: The specific coefficient values of $C(z;\theta)$ are uniquely determined from $\theta$ via the IO-approximation method and the discretization method satisfying Assumption 2.1.

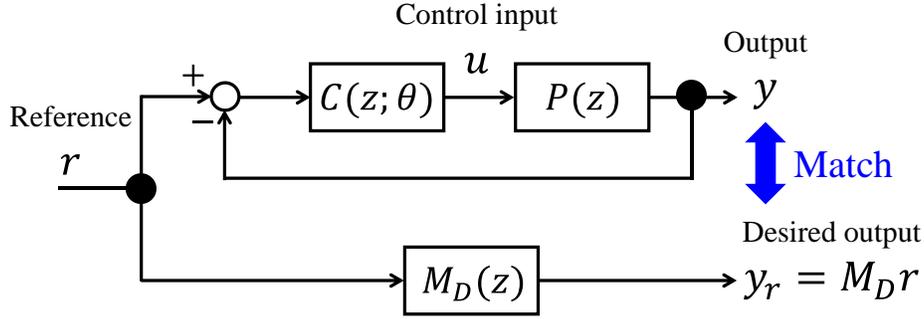

Fig. 1. Model-reference control problem.

## 3. Proposed approach

### 3.1. Fictitious reference signal

This section introduces the fictitious reference signal [42,53], which plays the central role in the proposed method. For given $u_{0,[0,N]}$, $y_{0,[0,N]}$, and $C(z;\theta)$, the fictitious reference $\tilde{r}_{[0,N]}(\theta) = \{\tilde{r}_k(\theta)\}_{k=0}^N$ is computed as follows [42]:

$$\tilde{r}_{[0,N]}(\theta) = \bigl(C(z;\theta)\bigr)^{-1} u_{0,[0,N]} + y_{0,[0,N]}, \tag{2}$$

i.e., $\tilde{r}_k(\theta) = \bigl(C(z;\theta)\bigr)^{-1} u_{0,k} + y_{0,k}$.

The fictitious reference signal has the following property [42]:

*Lemma 3.1*: $T(z;\theta)\tilde{r}_{[0,N]}(\theta) = y_{0,[0,N]}$ for any $\theta$ such that $1 + P(z)C(z;\theta) \neq 0$.

(*Proof*) See Eq. (8) in [42]. ∎

*Remark 3.1*: A previous study proposed FRIT, which is a direct data-driven approach of the model-reference problem described in Problem 2.1 [42]. However, if perfect model matching is not achievable by the controller class considered in design, the closed-loop system given by the tuning result may be unstable, as discussed in [54]. Several authors have proposed a modified version of FRIT, i.e., instability-detecting FRIT (IDFRIT), so as to avoid instability [54,55]. The fictitious reference signal is also employed in multi-model adaptive switching control with fine controller tuning [56] so as to tune the controller parameter (in [56], fictitious reference is called as *virtual*

*reference*). Note, in this method, that the closed-loop stability is guaranteed by the hysteresis switching logic (the HSL lemma) [56], which is different problem setup from this study.

### 3.2. IDFRIT with $\ell_1$-norm based model matching

The novel parameter tuning scheme is derived to solve Problem 2.1: IDFRIT with $\ell_1$-norm based model matching (L1-IDFRIT). L1-IDFRIT is the proposed approach in this study.

According to Lemma 3.1,

$$y_{0,k} = \sum_{\tau=0}^{k} t_\tau(\theta)\tilde{r}_{k-\tau}(\theta) \tag{3}$$

where the impulse response of $T(z;\theta)$ is denoted by $\{t_k(\theta)\}_{k=0}^\infty$. In matrix form, (3) becomes

$$[y_{0,[0,N]}] = \tilde{R}(\theta)t(\theta) \tag{4}$$

$$\tilde{R}(\theta) = \begin{bmatrix} \tilde{r}_0(\theta) & 0 & 0 & \ddots & 0 \\ \vdots & \tilde{r}_0(\theta) & 0 & \ddots & 0 \\ \tilde{r}_{N-2}(\theta) & \ddots & \tilde{r}_0(\theta) & 0 & \vdots \\ \tilde{r}_{N-1}(\theta) & \tilde{r}_{N-2}(\theta) & \ddots & \ddots & 0 \\ \tilde{r}_N(\theta) & \tilde{r}_{N-1}(\theta) & \tilde{r}_{N-2}(\theta) & \cdots & \tilde{r}_0(\theta) \end{bmatrix} \tag{5}$$

$$t(\theta) = [t_0(\theta) \quad t_1(\theta) \quad \cdots \quad t_N(\theta)]^T \tag{6}$$

where $[y_{0,[0,N]}] := [y_{0,0} \quad y_{0,1} \quad \cdots \quad y_{0,N}]^T$. Thus,

$$t(\theta) = (\tilde{R}(\theta))^{-1}[y_{0,[0,N]}]. \tag{7}$$

The output $y_{[0,N]}(\theta) = \{y_k(\theta)\}_{k=0}^N$ of the closed-loop system with $C(z;\theta)$ driven by $r_{0,[0,N]}$ can be computed as follows:

$$y(\theta) = R_0 t(\theta) \tag{8}$$

$$y(\theta) = [y_0(\theta) \quad y_1(\theta) \quad \cdots \quad y_N(\theta)]^T \tag{9}$$

$$R_0 = \begin{bmatrix} r_{0,0} & 0 & 0 & \ddots & 0 \\ \vdots & r_{0,0} & 0 & \ddots & 0 \\ r_{0,N-2} & \ddots & r_{0,0} & 0 & \vdots \\ r_{0,N-1} & r_{0,N-2} & \ddots & \ddots & 0 \\ r_{0,N} & r_{0,N-1} & r_{0,N-2} & \cdots & r_{0,0} \end{bmatrix}. \tag{10}$$

Consequently, the solution of Problem 2.1, i.e., the desired FOPID controller parameter, is obtained by solving the following optimization problem:

$$\text{minimize } J(\theta) \tag{11}$$

$$J(\theta) = \sum_{i=0}^{N} |y_i(\theta) - M_D(z)r_{0,i}| = \|\varepsilon(\theta)\|_1 \tag{12}$$

$$\varepsilon(\theta) = y(\theta) - R_0 m_D \tag{13}$$

$$m_D = [m_{D0} \quad m_{D1} \quad \cdots \quad m_{DN}]^T, \tag{14}$$

where the impulse response of $M_D$ at discrete time $k$ is denoted as $m_{Dk}$.

*Remark 3.2*: The loss function $J(\theta)$ of L1-IDFRIT, i.e., (12), explicitly reflects the information of the pole of the closed-loop system with $C(z;\theta)$ owing to (8) (note that each component of $t(\theta)$ is the impulse response of $T(z;\theta)$). This implies that if $J(\theta)$ is sufficiently small, the closed-loop system with $C(z;\theta)$ will be stable. Conversely, if $J(\theta)$ becomes a very large value during the optimization process, one can infer that such $\theta$ destabilizes the closed-loop system despite the boundedness of $r_{0,[0,N]}$ (Assumption 2.2). Therefore, L1-IDFRIT explicitly considers the BIBO stability of the closed-loop system from a bounded reference input to a bounded output in that the instability can be detected by $J(\theta)$ before implementing $C(z;\theta)$ to the real system. This feature yields a practically feasible controller parameter. Note that our approach has different theoretical background from multi-model adaptive switching control with fine controller tuning [56]: the stability of our approach relies on the preservation of the information of the closed-loop pole, whereas the approach in [56] guarantees the stability by the hysteresis switching logic.

*Remark 3.3*: The proposed tuning scheme, L1-IDFRIT, accepts various IO approximation methods and discretization methods as long as Assumption 2.1 holds. Therefore, L1-IDFRIT is applicable not only to the discrete-time FOPID controller but also to various types of discrete-time IO and FO controllers (e.g., the discrete-time IOPID controller) taking the transfer function. In addition, the same idea may be applicable to the continuous-time case.

### 3.3. Stability analysis in an asymptotic case

As discussed in Remark 3.2, L1-IDFRIT can explicitly consider the BIBO stability of the closed-loop system from a bounded reference input to a bounded output. Here, we further explore the stability of the closed-loop system tuned by the proposed method in the ideal case. Specifically, assume that $J(\theta) \to \alpha(< \infty)$ even for $N \to \infty$. In other words, (11) is successfully solved for infinite data points. Then, the following theorem holds:

*Theorem 3.1*: Let the solution of (11) be $\theta^*$. Then, $T(z;\theta^*)$ is BIBO stable.

(*Proof*) Owing to (13) and Assumption 2.2,

$$t(\theta) = R_0^{-1}\varepsilon(\theta) + m_D. \tag{15}$$

Therefore,

$$\|t(\theta)\|_1 = \left\|R_0^{-1}\varepsilon(\theta) + m_D\right\|_1 \leq \left\|R_0^{-1}\varepsilon(\theta)\right\|_1 + \|m_D\|_1 \leq |\gamma_{R0}|\|\varepsilon(\theta)\|_1 + \|m_D\|_1, \tag{16}$$

where $\gamma_{R0} \in \mathbb{R}_+$ represents the element of $R_0^{-1}$ having the maximum magnitude. Consequently, because $M_D(z)$ is BIBO stable, i.e., $\|m_D\|_1 < \infty$,

$$\lim_{N \to \infty} \|t(\theta)\|_1 = \alpha|\gamma_{R0}| + \|m_D\|_1 < \infty, \tag{17}$$

which shows the BIBO stability of $T(z; \theta^*)$ because of Lemma 2.1. ∎

*Remark 3.4*: The case considered above is a special case. Therefore, the condition that $J(\theta) < \infty$ as $N \to \infty$ is a sufficient condition and is conservative. Moreover, collecting infinite data points is impossible in practice. The derivation of practical conditions for BIBO stability based on finite data samples is left as a topic for future work. Note that, however, $J(\theta)$ is a good measure for inferring the stability of the resulting closed-loop system in practice, as discussed in Remark 3.2. Hence, L1-IDFRIT is a direct data-driven controller tuning scheme that yields a reliable FOPID controller in the ready-to-implement form.

### 3.4. Summary of the proposed FOPID controller tuning scheme

The proposed FOPID controller tuning procedure via L1-IDFRIT is summarized in Procedure 1 (if L1-IDFRIT is applied to the IO controllers, omit the procedures related to IO-approximation).

---

**Procedure 1**   Proposed tuning scheme (L1-IDFRIT)

1) Define $M_D(z)$. Determine the IO-approximation and the discretization methods.
2) Set $C(z; \theta_0) = \text{c2d}(\text{F2I}(C_{FOPID}(s; \theta_0)))$, where $\theta_0$ is the initial guess of the controller parameter.
3) Collect the data $u_{0,[0,N]}$ and $y_{0,[0,N]}$ via a closed-loop experiment using $C(z; \theta_0)$ and $r_{0,[0,N]}$.
4) Solve the optimization problem (11). The optimal solution of (11) is denoted as $\theta^*$.
5) Return $C(z; \theta^*) = \text{c2d}(\text{F2I}(C_{FOPID}(s; \theta^*)))$ as the resulting controller.

---

*Remark 3.5*: Since the optimization problem (11) is non-convex, step 4) in Procedure 1 may not be straightforward, especially if the problem dimension is large (e.g., a more complex fractional-order filter than the FOPID controller is chosen as the controller to be tuned). Moreover, this non-convexity makes it difficult to evaluate the globality of the solution, although local optima are still useful in practice. In the future, we will develop a convex relaxation technique [57] for the problem (11), since the convex relaxation significantly improves the tractability of the optimization problem. However, note that in our numerical examples shown in Section 4, the problem (11) was successfully solved by the particle swarm optimization algorithm implemented in the MATLAB. This fact implies that, at least for the tuning of the FOPID controller in practice, the problem (11) is so simple that it can be easily handled by standard solvers implemented in commercial numerical software.

## 4. Examples

Numerical examples are presented to demonstrate the validity of the proposed approach. In this study, the FOPID controller is IO-approximated using the Oustaloup recursive filter (order: 5; valid frequency range $(\omega_b, \omega_h) = (10^{-6}, 10^3)$ rad/s) [52][58]. The Tustin method is adopted for discretizing the plant model, the controller, and the reference model. Numerical computation of the IO approximation is performed using the FOMCON toolbox [59]. The optimization problem (11) of L1-IDFRIT is solved by the particle swarm optimization (PSO) algorithm. The PSO algorithm is implemented using the function `particleswarm` of the MATLAB Global Optimization Toolbox. Note that in all the examples, the plant model is used only for simulations for acquisition of the initial input/output data and for validation of the tuning result. The proposed tuning scheme does not require any plant model. Moreover, the initial data acquisition is performed only once; repeated data collection is unnecessary. The initial data are obtained via a closed-loop experiment in all the examples. Note that this study is not the naïve application of the PSO algorithm for tuning the FOPID controller but the development of the novel one-shot data-driven controller design technique. The PSO algorithm is merely the optimization solver for the new data-driven controller design scheme (Procedure 1).

### 4.1. Example #1: process model

The details of this example are summarized in Table 1. The plant model to be controlled is a process model, which is a challenging benchmark problem for evaluating the effectiveness of the control strategy [54,60]. Fig. 2 shows the data for the tuning scheme obtained by $C(z; \theta_0)$.

Table 2 summarizes the tuning result via the proposed strategy. The simulation result by $C(z; \theta^*)$ is shown in Fig. 3. In Fig. 3, the blue, black, and red lines correspond to the reference signal indicating the set point, the response of $M_D(z)$ (i.e., the reference model) for the set point, and that of $T(z; \theta^*)$. The black and red lines almost overlap. Table 2 and Fig. 3 confirm that the proposed tuning scheme yields the appropriate controller parameter exhibiting the desired closed-loop response.

Table 1. Conditions for Example #1.

| | |
|---|---|
| Sampling time [s] | 0.1 |
| Plant model $P(z)$ | $\text{c2d}(P(s)),\ P(s) = \dfrac{12s + 8}{20s^4 + 113s^3 + 147s^2 + 62s + 8}$ |
| Reference model $M_D(z)$ | $\text{c2d}(M_D(s)),\ M_D(s) = \left(\dfrac{1}{s+1}\right)^2$ |
| Controller parameters for initial experiment $\theta_0$ | $[K_{fp}\ \ K_{fi}\ \ \lambda\ \ K_{fd}\ \ \mu]^T = [1\ \ 0\ \ 1\ \ 0\ \ 1]^T$ |
| Parameter search range | $K_{fp}, K_{fi}, K_{fd} \in [0\ \ 10],\ \ \lambda, \mu \in [0\ \ 2]$ |
| Simulation time for initial data collection [s] | 100 |
| Reference signal for initial data collection $r_0$ | Unit step |

Table 2. Tuning results (Example #1).

| | |
|---|---|
| $J(\theta_0)$ | 496.1250 |
| $J(\theta^*)$ | 0.3805 |
| Tuned parameter $\theta^*$ | $[K_{fp}\ \ K_{fi}\ \ \lambda\ \ K_{fd}\ \ \mu]^T$ $= [2.7563\ \ 0.5105\ \ 0.9966\ \ 2.6412\ \ 0.8482]^T$ |

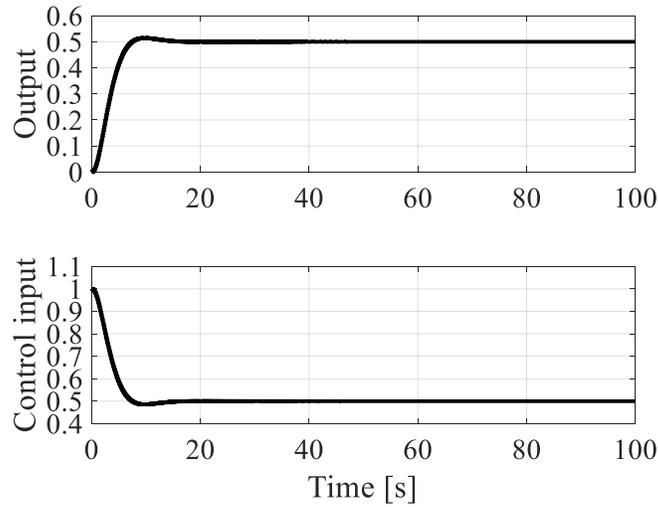

Fig. 2. Input/output time series data obtained by $C(z; \theta_0)$ (Example #1).

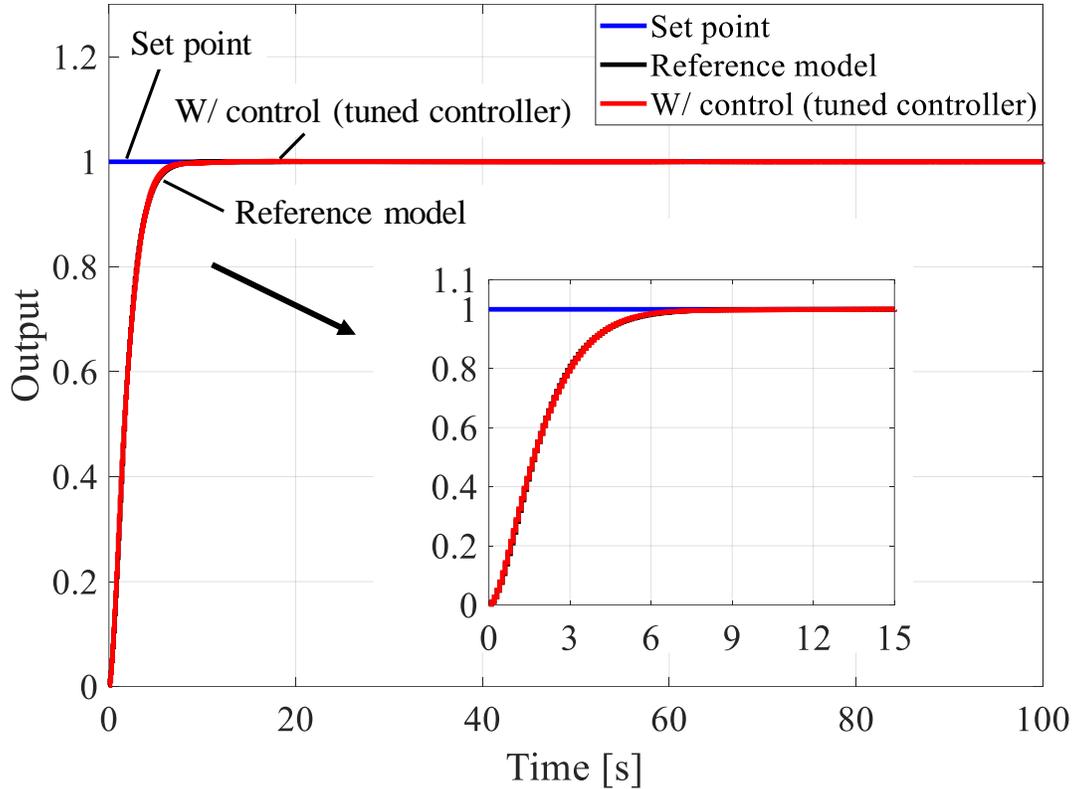

Fig. 3. Control result obtained by $C(z; \theta^*)$ (Example #1). The black and red lines almost overlap.

### 4.2. Example #2: process model with time delay

Table 3 summarizes the conditions for Example #2. In Example #2, the plant to be controlled is that in Example #1 with a time delay [54]. Meanwhile, the reference model does not include a time delay, implying that perfect model matching is impossible. Note that in such a situation, the resulting closed-loop system designed by conventional FRIT tends to be unstable, as shown in [54]. Fig. 4 describes the data for tuning obtained by $C(z; \theta_0)$.

Table 4 summarizes the tuning result via the proposed approach. Fig. 5 demonstrates the control result obtained by $C(z; \theta^*)$. The meaning of the blue, black, and red lines is the same as that in Fig. 3. As indicated by Table 4 and Fig. 5, the controller tuned by the proposed approach can control the plant without destabilization even if perfect model matching is not achievable.

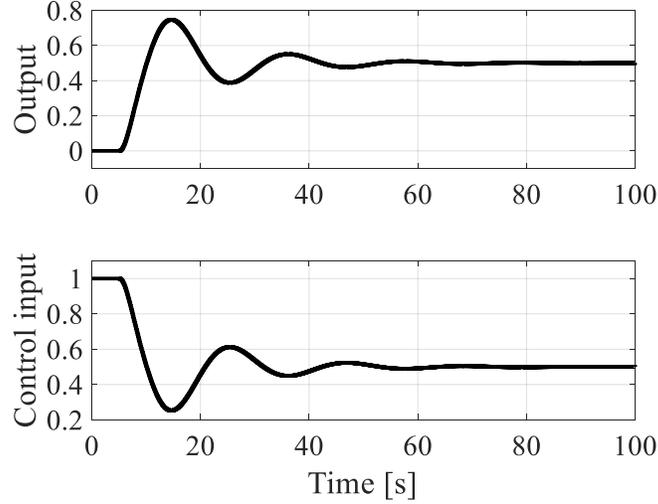

Fig. 4. Input/output time series data obtained by $C(z;\theta_0)$ (Example #2).

Table 3. Conditions for Example #2.

| | |
|---|---|
| Sampling time [s] | 0.1 |
| Plant model $P(z)$ | $\text{c2d}(P(s))$, $P(s) = \dfrac{12s+8}{20s^4+113s^3+147s^2+62s+8}e^{-5s}$ |
| Reference model $M_D(z)$ | $\text{c2d}(M_D(s))$, $M_D(s) = \left(\dfrac{1}{s+1}\right)^2$ |
| Controller parameters for initial experiment $\theta_0$ | $[K_{fp}\ \ K_{fi}\ \ \lambda\ \ K_{fd}\ \ \mu]^T = [1\ \ 0\ \ 1\ \ 0\ \ 1]^T$ |
| Parameter search range | $K_{fp}, K_{fi}, K_{fd} \in [0\ \ 10],\ \ \lambda, \mu \in [0\ \ 2]$ |
| Simulation time for initial data collection [s] | 100 |
| Reference signal for initial data collection $r_0$ | Unit step |

Table 4. Tuning results (Example #2).

| | |
|---|---|
| $J(\theta_0)$ | 508.6346 |
| $J(\theta^*)$ | 53.3317 |
| Tuned parameter $\theta^*$ | $[K_{fp}\ \ K_{fi}\ \ \lambda\ \ K_{fd}\ \ \mu]^T$ $= [1.4675\ \ 0.1368\ \ 1.0147\ \ 5.0724\ \ 1.3177]^T$ |

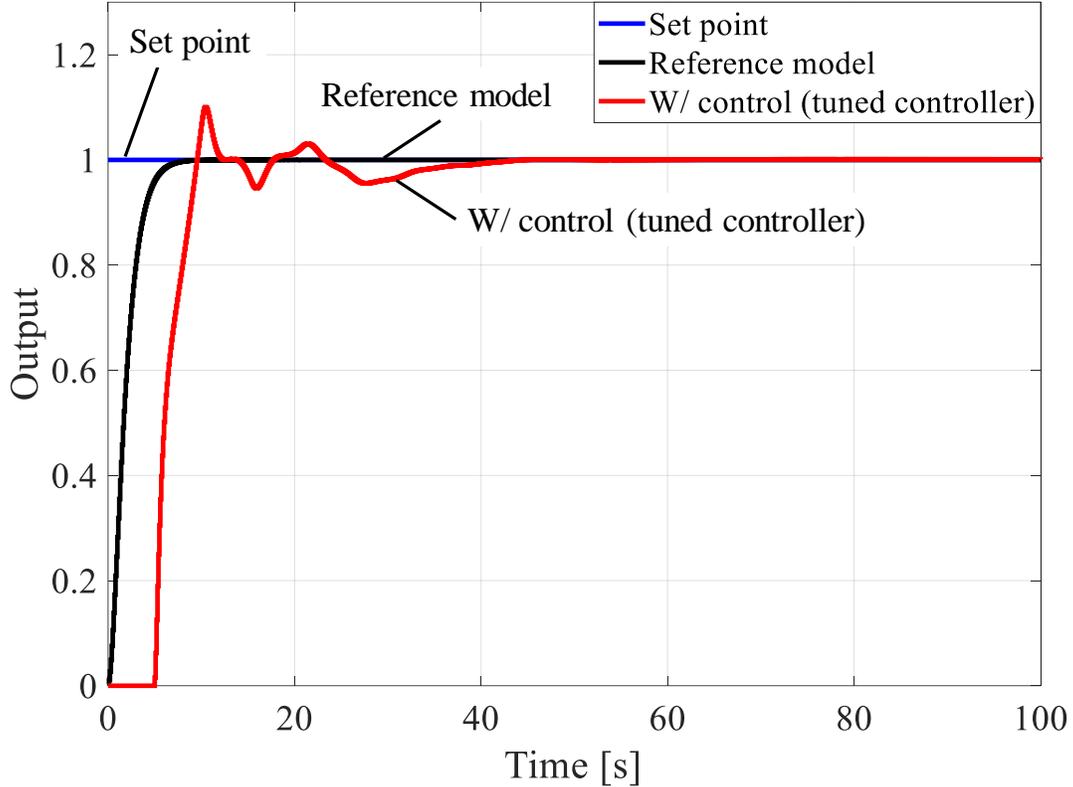

Fig. 5. Control result obtained by $C(z; \theta^*)$ (Example #2).

### 4.3. Example #3: flexible transmission (IOPID vs. FOPID)

In this example, the plant to be controlled is a flexible transmission model [40], which is often referred to as a challenging benchmark problem of digital control design. Moreover, this example shows the applicability of the proposed method to the conventional IOPID controller and compares the IOPID and FOPID controllers from the viewpoint of control performance. Specifically, the discrete-time IOPID controller $C_{IO}(z; \theta)$ takes the following form in this study, where the discretization method is the Tustin method:

$$C_{IO}(z; \theta) = \text{c2d}(C_{PID}(s; \theta)) \tag{18}$$

$$C_{IOPID}(s; \theta) = K_p + K_i s^{-1} + K_d s. \tag{19}$$

Note that (19) can be regarded as a special case of (1), where $\lambda = \mu = 1$. Hereafter, the discretized FOPID controller with controller parameter $\theta$ is denoted as $C_{FO}(z; \theta)$.

Table 5 summarizes the details of this example. The simulation time for initial data collection is 4 s, and the sampling time is 0.05 s. Note that the initial controllers $C_{FO}(z; \theta_0)$ and $C_{IO}(z; \theta_0)$ are identical since both $\lambda$ and $\mu$ of the initial controller $C_{FO}$ are 1. Thus, the initial input and output data for tuning $C_{FO}$ is also the same as that for tuning $C_{IO}$. Fig. 6 shows the initial input/output data obtained by $C_{FO}(z; \theta_0)(= C_{IO}(z; \theta_0))$. The initial controller seems to destabilize the closed-loop system although the obtained initial data does not take extreme values.

Table 6 summarizes the tuning results by the proposed method. The initial loss function value $J(\theta_0)$ via $C_{FO}(z;\theta_0)$ is the same as that via $C_{IO}(z;\theta_0)$ because $C_{FO}(z;\theta_0) = C_{IO}(z;\theta_0)$ as mentioned above. Fig. 7 shows the control result of the FOPID and IOPID controllers tuned by the proposed approach. In Fig. 7(a), the blue, black, green, and red lines show the setpoint reference, the response of the reference model, the control result by the tuned FOPID controller, and the control result by the IOPID controller, respectively. Fig. 7(b) plots the time history of the absolute value of the tracking errors, i.e., $|T_{IO}(z;\theta^*)r_0 - M_D(z)r_0|$ and $|T_{FO}(z;\theta^*)r_0 - M_D(z)r_0|$, where $T_{IO}(z;\theta^*)$ and $T_{FO}(z;\theta^*)$ are the closed-loop transfer functions from the reference signal to the output using $C_{IO}(z;\theta^*)$ and $C_{FO}(z;\theta^*)$, respectively. The green and red lines in Fig. 7(b) shows those of $C_{IO}(z;\theta^*)$ and $C_{FO}(z;\theta^*)$, respectively. Fig. 7(c) demonstrates the time history of the control input values: the green and red lines correspond to $C_{IO}(z;\theta^*)$ and $C_{FO}(z;\theta^*)$, respectively. Table 5 and Fig. 7 clearly demonstrate that the value of $J(\theta)$ is successfully reduced and the resulting FOPID and IOPID controller realizes tracking control. This result confirms the applicability of the proposed approach, L1-IDFRIT, to the tuning of the IO controllers. Moreover, Fig. 7 also shows that the FOPID controller outperforms the IOPID controller from the viewpoint of both control performance and efficiency of the control action. Such an efficient control action of the FOPID controller is owing to the long-memory effect of the FO integration and differentiation [61]. The reduction of control input is one significant advantage of FO control over IO control [12,62].

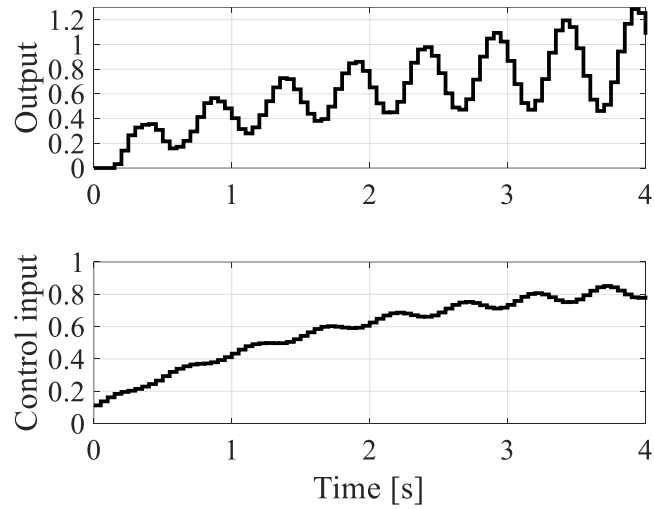

Fig. 6. Input/output time series data obtained by $C(z;\theta_0)$ (Example #3).

Table 5. Conditions for Example #3.

| | | |
|---|---|---|
| Sampling time [s] | | 0.05 |
| Plant model $P(z)$ | | $\dfrac{z^{-3}(0.28261 + 0.50666z^{-1})}{1 - 1.41833z^{-1} + 1.58939z^{-2} - 1.31608z^{-3} + 0.88642z^{-4}}$ |
| Reference model $M_D(z)$ | | $M_D(z) = \dfrac{z^{-3}(1-\alpha)^2}{(1-\alpha z^{-1})^2}, \quad \alpha = e^{-0.05\bar{\omega}}, \quad \bar{\omega} = 10$ |
| Controller parameters for initial experiment $\theta_0$ | $C_{IO}$ | $[K_p \quad K_i \quad K_d]^T = [1 \times 10^{-1} \quad 5 \times 10^{-1} \quad 0]^T$ |
| | $C_{FO}$ | $[K_{fp} \quad K_{fi} \quad \lambda \quad K_{fd} \quad \mu]^T = [1 \times 10^{-1} \quad 5 \times 10^{-1} \quad 1 \quad 0 \quad 1]^T$ |
| Parameter search range | $C_{IO}$ | $K_p, K_i, K_d \in [0 \quad 5]$ |
| | $C_{FO}$ | $K_{fp}, K_{fi}, K_{fd} \in [0 \quad 5], \quad \lambda, \mu \in [0 \quad 2]$ |
| Simulation time for initial data collection [s] | | 4 |
| Reference signal for initial data collection $r_0$ | | Unit step |

Table 6. Tuning results (Example #3).

| | | |
|---|---|---|
| $J(\theta_0)$ | | 28.6451 |
| $J(\theta^*)$ | $C_{IO}$ | 1.1129 |
| | $C_{FO}$ | 0.8087 |
| Tuned parameter $\theta^*$ | $C_{IO}$ | $[K_p \quad K_i \quad K_d]^T = [0.0214 \quad 3.3025 \quad 0.0209]^T$ |
| | $C_{FO}$ | $[K_{fp} \quad K_{fi} \quad \lambda \quad K_{fd} \quad \mu]^T = [1.0894 \times 10^{-9} \quad 3.3490 \quad 1.0018 \quad 0.0242 \quad 0.9448]^T$ |

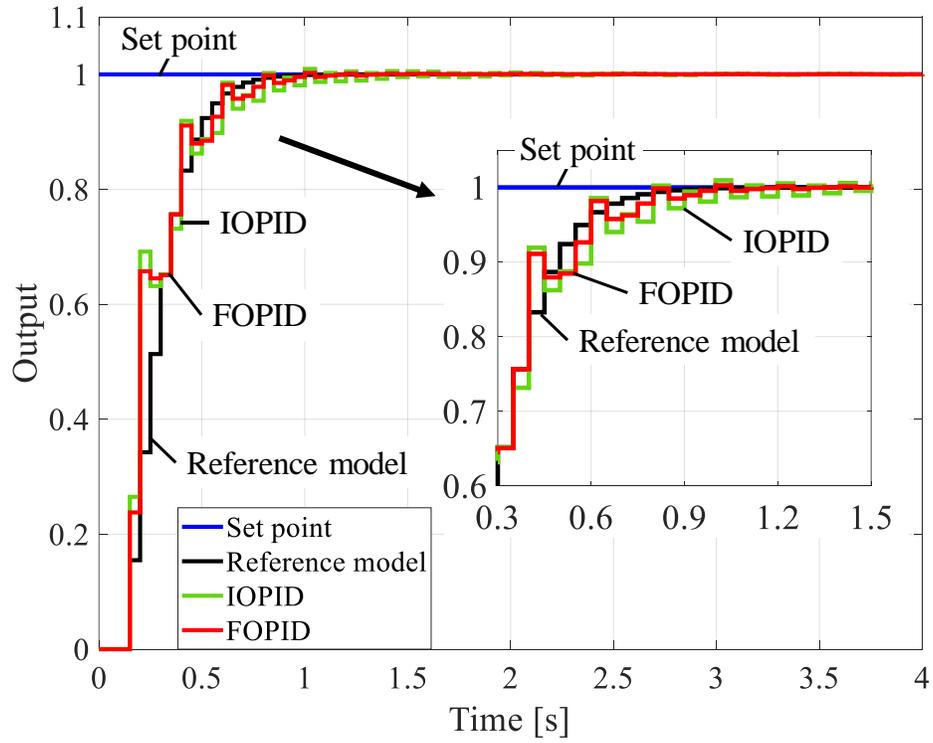

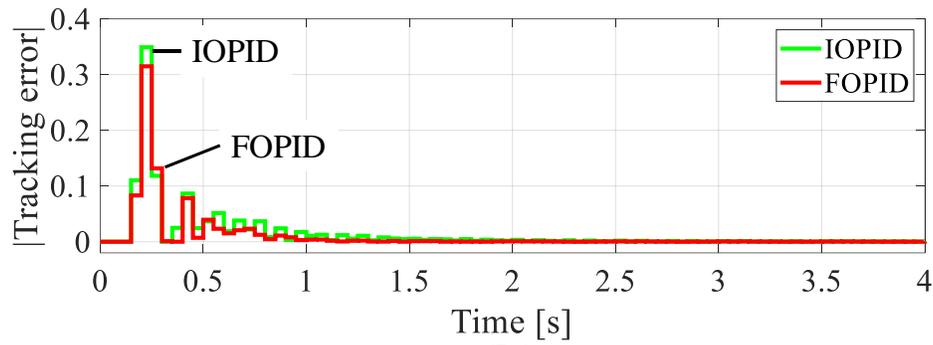

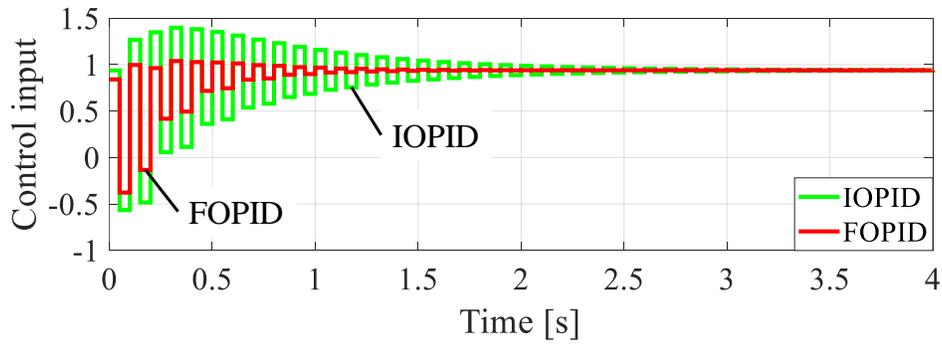

Fig. 7. Control results obtained by $C_{IO}(z;\theta^*)$ and $C_{FO}(z;\theta^*)$ (Example #4): (a) time history of the closed-loop response; (b) absolute value of the tracking error; (c) control input.

## 5. Discussion

As demonstrated in Examples #1–#3, the proposed tuning technique successfully provides appropriate parameters for the FOPID and IOPID controllers. Specifically, the parameter values, which minimize the value of $J(\theta)$ as shown in Tables 2, 4, and 6 yield the desired closed-loop response. Moreover, the resulting closed-loop system exhibits no destabilization even when perfect model matching is impossible, as demonstrated in Example #2. Such stability is attributed to the explicit consideration of the BIBO stability by the proposed loss function $J(\theta)$ shown in (12). Accordingly, these results confirm the effectiveness of the proposed method.

The examples presented in Section 4 involve various situations and controlled objects: process control where model matching may be achievable (Example #1), process control where model matching is impossible (Example #2), and a flexible transmission system (Example #3). These examples demonstrate the applicability of the proposed technique to various practical control problems. The proposed method is well suited for process control possibly having a time delay, as shown in Examples #1 and #2. Example #3 highlights the applicability of the proposed method to mechanical systems. Example #3 also shows that the proposed method can be used for tuning not only the FO controller but also IO controllers.

In the direct data-driven controller tuning for the model-reference control problem, a major challenge is the design of the reference model. An inappropriate reference model results in inadequate model matching, leading to destabilization of the resulting closed-loop system. For instance, the conventional FRIT returns a destabilizing controller even though the objective function value is small [54]. On the other hand, Example #2 demonstrates the validity of the proposed method in the situation where perfect model matching is inherently unattainable owing to the property of the plant and the reference model. This result confirms the reliability of the proposed method in terms of the closed-loop stability. Optimizing not only the controller parameter but also the reference model will enhance the reliability.

The proposed method requires only one-shot input/output data. In general, experimental data is corrupted by noise. Although the noise may affect the performance of data-driven control, many studies have examined strategies for addressing noise, such as regularization [37], total variation denoising [54], discrete Fourier transform via periodization [55], and reformulation of the optimization problem using slack variables [63]. In the future, the noise tolerance of the proposed method will be evaluated and an appropriate handling method for noise will be explored.

This study has established a simple and practical tuning scheme for the FOPID controller. Despite the fact that FO control including FOPID has clear advantages over conventional IO control as demonstrated in Example #3, the majority of control systems in practice involve IO controllers such as the IOPID controller [12]. One study has pointed out that a simple tuning rule, which can specify the required performance and robustness, and reliable IO approximation are necessary for industrial application of FOPID control [12]. The proposed method is an effective solution satisfying both requirements. Specifically, only one-shot input/output data is needed for the proposed tuning scheme and the plant model is not required. The proposed method allows one to specify the desired closed-loop property by giving an appropriate reference model. During the tuning, the closed-loop stability and performance can be explicitly evaluated for the IO-approximated and discretized (i.e., *ready-to-implement*) form of the FOPID controller. Moreover, the proposed tuning algorithm accepts not only the FOPID controller but also various FO (and IO) controllers in the transfer function form as discussed in Remark 3.3. Consequently, this study can facilitate industrial application of FO control by providing a practical implementation technique. Spread of FO control should significantly improve the safety, reliability, and performance of many practical automatic control systems.

## 6. Conclusion

This study presented a simple and practical tuning scheme (L1-IDFRIT) for the FOPID controller. The proposed tuning method needs only one-shot input/output data, and the plant model is not required. The tuning problem is defined as the model-reference control problem. The loss function is derived using a fictitious reference signal, which can be computed using the input/output data. The loss function value explicitly considers the BIBO stability of the resultant closed-loop system from a bounded reference input to a bounded output. The model-reference control problem was reformulated as the minimization of the loss function. The validity of the proposed method was examined via numerical simulations. As a result, the proposed tuning scheme enables the FOPID controller to achieve good control performance and closed-loop stability even if the perfect model-matching is unachievable. The proposed approach was shown to be applicable not only to the FOPID controller but also to IOPID controllers. In addition, the comparative study demonstrated that the FOPID controller can outperform the IOPID controller from the viewpoint of control performance and reduction of control effort.

In the future, the effectiveness of the proposed technique will be experimentally evaluated using real-world plants. Furthermore, a suitable optimization algorithm for L1-IDFRIT will be explored.